# Limitations on Sub-Diffraction Imaging with a Negative Refractive Index Slab

David R. Smith*, David Schurig, Marshall Rosenbluth, Sheldon Schultz,
*Department of Physics, University of California, San Diego, 9500 Gilman Drive, La Jolla, CA 92093-0319*

S. Anantha Ramakrishna, John B. Pendry
*The Blackett Laboratory, Imperial College, Prince Consort Road, London, SW7 2BZ*

Recently it has been proposed that a planar slab of material, for which both the permittivity and permeability have the values of –1, could bring not only the propagating fields associated with a source to a focus, but could also refocus the nonpropagating near-fields, thereby achieving a subdiffraction image. In this work we discuss the sensitivity of the subwavelength focus to various slab parameters, pointing out the connection to slab plasmon modes. We also note and resolve a paradox associated with the perfect imaging of a point source. We conclude that subwavelength resolution is achievable with available technology, but only by implementation of a critical set of design parameters.

*Corresponding author: drs@sdss.ucsd.edu

In 1964 Veselago suggested that if a material could be found in which the permittivity (ε) and permeability (μ) were both simultaneously negative, it should be regarded as having a negative index-of-refraction [1]. Veselago termed such a material as *left-handed*, and predicted that reversals of a number of fundamental electromagnetic phenomena would occur as a consequence of the change of sign in the refractive index. As an example, Snell's law requires that electromagnetic waves incident upon the interface between two materials having opposite signs of refractive index be refracted to the *same* side of the normal, rather than to the opposite side as for all other prior known materials. In considering the impact of this modification of Snell's law, Veselago further pointed out that for a particular set of material properties, the rays from an electromagnetic source placed near a left-handed planar slab would be refocused, first inside and again outside the slab.

Pendry recently extended Veselago's analysis [2], showing that in addition to the far-field components associated with a source being brought to a focus by the slab, the nonpropagating near-field components could also be recovered in the image. It was therefore proposed that the image of the source created by a planar slab could, in principle, contain all of the information associated with the source object, thereby achieving resolution well beyond that of the diffraction limit. For this reason the slab was described as a *perfect lens*, although the slab does not focus rays from infinity. We nevertheless maintain this description here, referring specifically to a planar slab with μ=−1 and ε=−1 (no losses) as a perfect lens.

The subwavelength imaging associated with the perfect lens was a surprising result, stimulated by the first experimental demonstration [3] of a left-handed material (LHM), in which ε<0 and μ<0. However, far from being a continuous material, the experimental LHM was comprised of two interlaced periodic arrays of copper elements, one array being composed of split ring resonators (SRRs) [4], and the other, wires [5]. The response of the SRR array gave rise to an effective magnetic permeability, while the response of the wires gave rise to an effective electric permittivity. Each of the arrays considered separately had frequency bands over which either the permittivity or the permeability was negative; when the arrays were combined into a single medium, the resulting composite exhibited a region over which *both $\varepsilon$ and $\mu$ were negative*. Such a composite structure when probed by wavelengths larger than the unit cell size has been termed a *metamaterial* [6].

Recently, a two-dimensionally isotropic LHM [7] was used to demonstrate explicitly the property of negative refraction [8] at microwave frequencies. This experiment confirmed the fundamental concept behind the perfect lens, and provides experimental evidence that a planar LHM will, in fact, refocus the far field radiation of a point source. However, the values of the electromagnetic parameters, and spatial periodicity, of this metamaterial sample are significantly different than those specifying the perfect lens. The immediate question, and subject of this paper, is whether or not the subwavelength focusing can be observed using any practically obtained or fabricated material. The answer to this question will depend on a number of important factors, most important being the degree of subwavelength focusing desired.

To perform the analysis here, we follow the procedure in [2], expressing the field from a radiating source as a sum over propagating and nonpropagating plane waves. For the purposes of the present calculations, we will restrict ourselves to a two-dimensional line source with the electric field having S-polarization, so that the expression for the field has the form

$$\mathbf{E}(x,z,t) = \sum_{k_x} \mathbf{E}(k_x) \exp(ik_z z + ik_x x - i\omega t), \qquad (1)$$

$k_z$ and $k_x$ are the components of the wave vector normal to and parallel to the slab, respectively. Outside the slab, the homogeneous wave equation must be satisfied, leading to a dispersion relation relating the frequency $\omega$ and the components of the wave vector, or

$$k_z = \sqrt{\frac{\omega^2}{c^2} - k_x^2}. \tag{2}$$

The general expansion of the source fields contains both propagating and nonpropagating plane wave terms. Those modes for which $k_x < \omega/c$ are propagating, while those for which $|k_x| > \omega/c$ evanescently decay along the propagation direction (z). Ordinary (positive index) optical elements can refocus the propagating components, but the exponentially decaying inhomogeneous components are always lost, leading to the diffraction limit for focusing to an image [2].

In principle, there always exists an exponentially growing solution for an electromagnetic wave in any planar slab geometry, in addition to the exponentially decaying solution; however, this growing solution is never the dominant one in systems with positive refractive index. It is only in the presence of negative $\varepsilon$ or $\mu$ that there exists the possibility that the dominant decaying solution can be cancelled, leaving only the growing solution. To see how this can come about, we start with the expression for the transmission coefficient for an S-polarized plane wave incident on a slab of thickness d and arbitrary values of $\varepsilon$ and $\mu$, or

$$t_s^{-1} = \cos(q_z d) - \frac{i}{2}\left(\frac{\mu k_z}{q_z} + \frac{q_z}{\mu k_z}\right)\sin(q_z d), \tag{3}$$

where $q_z$ is the z-component of the propagation vector within the slab, and has the value

$$q_z = \sqrt{\varepsilon\mu\frac{\omega^2}{c^2} - k_x^2}. \tag{4}$$

We have absorbed the phase factor exp(-$k_z$d) into the definition of the transmission coefficient, as this factor does not affect the results presented here. The expression for P-polarized waves is similar to Eq. 3, with the explicitly appearing $\mu$ replaced by $\varepsilon$. We apply all of our arguments to the S-polarization terms in this paper, as the results for P-polarization follow trivially.

As we are primarily concerned with nonpropagating modes, we assume that $k_x > \sqrt{\varepsilon\mu}\,\omega/c$ and $k_x > \omega/c$, so that $q_z \to iq_z$ and $k_z \to ik_z$. The transmission coefficient then has the form

$$t_S = \left\{ e^{q_z d}\left[\frac{1}{2} + \frac{1}{4}\left(\frac{\mu k_z}{q_z} + \frac{q_z}{\mu k_z}\right)\right] + e^{-q_z d}\left[\frac{1}{2} - \frac{1}{4}\left(\frac{\mu k_z}{q_z} + \frac{q_z}{\mu k_z}\right)\right] \right\}^{-1}. \tag{5}$$

where $q_z$ and $k_z$ are real.

When $\mu$ is positive, the first term in brackets will always dominate the behavior of the transmitted wave for large d, and an evanescent component will decay exponentially through the slab. The only condition under which the normally dominant solution is *entirely absent* occurs when both ε and μ are equal to –1. Under this condition, $t_s = \exp(+q_z d)$, and every evanescent component from the source field will grow exponentially, thus exactly reproducing the source field in the image plane. But the balance is delicate; any deviation from the perfect lens condition, however small, will result in an imperfect image that degrades exponentially with slab thickness d, until the usual diffraction limit is reached.

The effect of the various parameters on the image for a given slab thickness can be estimated by determining those values of parameters which cause the two terms in Eq. (5) in brackets to be roughly equivalent. Before performing this analysis, we first note that the denominator of the transmission coefficient will generally have zeros. After some minor algebra, we find that when the denominator vanishes the following dispersion formulas hold:

$$\tanh\left(\frac{q_z d}{2}\right) = -\frac{\mu k_z}{q_z} \tag{6}$$

and

$$\coth\left(\frac{q_z d}{2}\right) = -\frac{\mu k_z}{q_z}, \tag{7}$$

which correspond to the dispersion relations for slab plasmon polaritons [9,10]. When either of these dispersion relations is satisfied, resonant surface modes can exist on the slab for certain values of $k_x$ in the absence of any source field. The direct excitation of these surface modes for imaging applications is undesirable, as these resonances will be disproportionately represented in the image; yet, the existence of these resonances is essential, as the recovery of the evanescent modes can be seen as the result of driving the surface plasmon far off-resonance. Note that it is only when ε and μ are both equal to −1 that *no surface mode can be resonantly excited for any finite value of $k_x$*. Since this will never be exactly true (radiation damping or thermal noise, for example, will introduce some degree of loss or limitation at every frequency), the effect of these resonances must always be considered in the operation of a practical imaging configuration. The appearance of resonances can be seen in Fig. (1), which shows the transfer function from source to image as a function of $k_x$.

The perfect lens solution in the absence of losses is very special, and can lead to paradoxical interpretations. For example, a point source placed on one side of the slab apparently produces a singularity in the image plane, in which fields propagate toward the singularity from one half-space, and emerge from the singularity into the other half-space. While the source singularity corresponds to a solution allowed by the wave equation, the two singularities that would be produced by an ideal lossless perfect lens would not represent valid solutions. To resolve the apparent inconsistencies, an initial value calculation was performed using Laplace Transform methods (details to be presented elsewhere). This calculation showed that the solution was causal, that a time delay was necessary to establish the negative group velocity waves in the slab, and that transient waves occurred near the roots of Eqs. 6 and 7. The transients die away in a dissipation time, leaving a steady-state solution whose transfer function Fourier components approach those given by Eq. 5, with the exception of some algebraically ($t^{-1.5}$) decaying terms. These latter terms depend on $k_x$, and in particular approach their asymptotic limit only for times proportional to $k_x$. In the absence of losses, the image of a point source is thus never achieved at finite times, a behavior that has been recently observed in finite-difference time-domain studies of focusing by a negative-refractive index slab [11]. With dissipation this limitation exists even for the asymptotic solution, as will be discussed below.

If the real parts of either $\varepsilon$ or $\mu$ deviate from $-1$, one or more surface modes will be excited by the source, and the range of values of $k_x$ for which the evanescent modes are growing will be reduced. This can be seen in Fig. (1), where $\mu$ is varied from $-1$. For small deviations from the perfect lens condition we can find an approximate expression for the resolution of the slab. Assuming that $\varepsilon = -1$ and $\mu = -1 + \delta\mu$, the two terms in the denominator of Eq. 5 are of approximately the same magnitude when

$$2k_z d = -\ln\left|(\delta\mu)^2\right|. \tag{8}$$

We have assumed that the limit occurs when the value of $k_x$ is large; in this case, $|q_z| \sim |k_z| \sim |k_x|$.

A reasonable definition of enhanced resolution is the minimum feature resolved (or recovered) in the image relative to the excitation wavelength. At the maximum $k_x$ as determined from Eq. (8), this minimum feature will be on the scale of $\lambda_{min} = 2\pi/k_x$. Applying this definition, we arrive at the following formula for the enhancement of the resolution, R, of the lens as a function of small deviations in the permeability, or

$$R \equiv \frac{\lambda}{\lambda_{min}} = -\frac{\ln|\delta\mu|}{2\pi}\frac{\lambda}{d}. \tag{9}$$

The validity of this approximate expression for the resolution enhancement can be seen by comparing Eq. 9 with Figure 1. For example for a deviation of $\delta\mu=0.005$, the numerically computed transfer function shown in Figure 1 shows that the range of $k_x$ values near unity is approximately up to $k_x \sim 8k_0$, while Eq. 9 predicts R~8.4.  The comparison is, of course, complicated by the plasmon resonances that terminate the transfer function.

The dependence of the resolution enhancement R on the deviation from the perfect lens condition is critical.  The ratio $\lambda/d$ (wavelength to slab length) dominates the resolution, the logarithm term being a relatively weakly varying function.  For example, if $\lambda/d=1.5$, we find that to achieve an R of 10, $\delta\mu$ must be no greater than $\sim 6 \times 10^{-19}$!  However, for $\lambda/d=10$, $\delta\mu$ can vary by as much as ~0.002 to achieve the same resolution enhancement.  Similar deviations in either the real or the imaginary part of the permeability will result in the same resolution [data not shown], although the surface resonances are much stronger when the deviation occurs in the real part.  Note that when the losses are on the order of the deviation in the real part of the permeability, the resonances are significantly reduced.  Although the inherent resolution is not improved, the associated image should be closer to the source field distribution.

The effect of varying $\varepsilon$ (for S-polarization) is much smaller than that associated with $\mu$, and we do not discuss the effects of this variation further.

Because no naturally occurring material possesses simultaneously a negative permittivity and negative permeability, especially at lower frequencies (e.g., gigahertz), the most likely candidate structure to produce a left-handed focusing slab is a structured metamaterial.  While not necessarily required, a periodic ordering of subwavelength scattering elements has been a characteristic feature of metamaterials demonstrated thus far.  While the effect of periodicity is minimal for propagating plane waves whose wavelength is much larger than the repeated unit cell size, periodicity produces a significant effect on the nonpropagating components having large transverse wave number. This places an additional important limitation on the resolution achievable with a metamaterial slab.

We can introduce periodicity into this analysis in an approximate manner. The wave equation in a medium, assuming that **E** is polarized in the y-direction, is

$$\frac{\partial^2 E_y}{\partial x^2} + \frac{\omega^2}{c^2} n^2(x) E_y = q_z^2 E_y. \tag{10}$$

We have assumed that μ does not vary spatially, and that the periodic variation in ε(x) is only in the transverse (x-) direction. Under these assumptions, $n^2(x)=\varepsilon(x)\mu= n^2(x+a)$, where a is the repeat distance.

Due to the periodicity in $n^2(x)$, Eq. (10) in Fourier space will involve sums over reciprocal lattice vectors $g_n=2n\pi/a$. We may thus write

$$n^2(x) = \sum_g n_g^2 e^{igx} \tag{11}$$

and

$$E_y(x) = \sum_g E_g e^{i(k_x+g)x}. \tag{12}$$

Using Eqs. (11) and (12) in Eq. (10), and integrating over a unit cell, we arrive at the following eigenequation

$$\sum_{g'} \left[ \frac{\omega^2}{c^2} n^2_{g-g'} - |k_x + g|^2 \delta_{g,g'} \right] E_{k_x,g'} = q_z^2 E_{k_x,g} \tag{13}$$

For every value of ω and $k_x$, there exists a set of solutions of Eq. (13) indexed by $g_n=2\pi n/a$.

In the absence of periodicity, Eq. (13) reduces to the usual dispersion relation, Eq. (2). For values of $k_x$ smaller than ω/c, the eigenvalues $q_z$ are real, while for values of $k_x$ greater than ω/c the eigenvalues $q_z$ are imaginary, as indicated by the solid and dashed curves in Fig. 2. Since we assume a<<λ, the introduction of a periodic modulation introduces gaps into the imaginary branches, also shown in Fig. 2.

To obtain a rough estimate of the limitation that periodicity imposes on the resolution enhancement, we solve Eq. (13) for an index having the form

$$n^2(x) = 1 + 2\Delta \sin \frac{2\pi x}{a} \tag{14}$$

Using this form in Eq. (13), and assuming the periodic modulation is sufficiently weak that only two bands need be considered, we find the modified dispersion relation can be found by evaluating

$$\begin{vmatrix} k_0^2 - k_x^2 - q_z^2 & k_0^2 \Delta^2 \\ k_0^2 \Delta^2 & k_0^2 - (k_x - g)^2 - q_z^2 \end{vmatrix} = 0, \tag{15}$$

where $g=2\pi/a$ and $k_0=\omega/c$. In the case that $\Delta<<1$, the bands only weakly couple. Ignoring the effects of higher mode excitation in the transmission calculation, we may retain Eq. (3) for the transmission coefficient, and consider the periodicity as only modifying the dispersion characteristics of the lower branch. Solving for the roots of Eq. (15), we find

$$q_z^2 - k_z^2 = \frac{-(g^2 - gk_x) + \sqrt{(g^2 - gk_x)^2 + (2k_0\Delta^2)^2}}{2} \tag{16}$$

where $k_z$ is the z-component of the propagation vector in free space. When the two terms in the denominator of Eq. (5) are roughly equal, we have the condition

$$q_z d = -\frac{1}{2}\ln\left|\frac{(q_z - k_z)^2}{(q_z + k_z)^2}\right|; \quad -\ln\left|\frac{1}{4}\frac{(q_z^2 - k_z^2)}{k_z^2}\right| \tag{17}$$

Substituting Eq. (16) into the above yields

$$q_z d = -\ln\left|\frac{1}{2}\frac{k_0^2 \Delta^2}{k_z^2}\right| + \sinh^{-1}\left|\frac{g^2 - k_x^2}{2k_0^2 \Delta^2}\right|. \tag{18}$$

In the limit $k_0\Delta/g<<1$, and using the definition of resolution in Eq. (9), we find the resolution enhancement to be

$$R \equiv \frac{\lambda}{\lambda_{min}} = \frac{1}{2\pi}\ln\left(\frac{\lambda^2}{a^2\Delta^4}\right)\frac{\lambda}{d}. \tag{19}$$

A similar resolution enhancement limit to that obtained by a variation in material parameter is thus imposed by the introduction of periodic modulation. Similar to Eq. (9), the effects of periodicity enter logarithmically, and again lead to a critical dependence. As an example, an R of ten for a slab with $\lambda/d=10$ and $\Delta\sim1$ requires a periodicity of $\sim\lambda/20$ .

It should be noted that we have applied a definition for resolution in this work that may not always be appropriate for all applications. For instance, the effect of the excitation of surface plasmons leads to a broad background in the image plane, with the subwavelength features superposed [12]. However, for certain applications, the subwavelength information obtained from such an image may be of value.

Our conclusion from this analysis indicates that the perfect lens effect exists for a fairly restricted region of parameter space. Yet, demanding as these specifications are, achieving subwavelength resolution is possible with current technologies. Negative refractive indices have been demonstrated in structured metamaterials. Such materials can be engineered to have tunable material parameters, and so achieve the optimal conditions. Losses can be minimized in structures utilizing superconducting elements, or in which active elements can be used. Finally, consideration of alternate structures may lead to configurations of left-handed materials that can achieve subwavelength focusing, but with reduced constraints. We are currently studying such structures, and will present those results in a separate publication [12].

This work was supported under by DARPA through a grant from ONR (Contract No. N00014-00-1-0632) and AFOSR (Grant No. F49620-01-1-0440). We thank Dr. P. M. Platzman (Lucent) for valuable discussions.

REFERENCES:

1. V. G. Veselago, *Soviet Physics USPEKHI* **10**, 509 (1968).

2. J. B. Pendry, *Phys. Rev. Lett.* **85**, 3966 (2000).

3. D. R. Smith, W. Padilla, D. C. Vier, S. C. Nemat-Nasser, S. Schultz, *Phys. Rev. Lett.* **84**, 4184 (2000).

4. J. B. Pendry, A. J. Holden, D. J. Robbins, W. J. Stewart, *IEEE Trans. MTT*, **47**, 2075 (1999).

5. J. B. Pendry, A. J. Holden, W. J. Stewart, I. Youngs, *Phys. Rev. Lett.* **76**, 4773 (1996).

6. R. Walser, *private communication.*

7. R. A. Shelby, D. R. Smith, S. C. Nemat-Nasser, and S. Schultz, *Appl. Phys. Lett.* **78**, 489 (2001).

8. R. A. Shelby, D. R. Smith, S. Schultz, *Science* **292**, 77 (2001).

9. R. Ruppin, *J. Phys.: Condens. Matter* **13**, 1811 (2001).

10. H. Raether, *Surface Plasmons* (Springer-Verlag, Berlin, 1988).

11. R. Ziolkowski and E. Heyman, *Phys. Rev. E*, **64**, 056625 (2001).

12. S. A. Ramakrishnan, J. B. Pendry, D. Schurig, D. R. Smith, S. Schultz, *Journal of Modern Optics*, in press (2002).

FIGURE CAPTIONS

**Figure 1:** Transfer function [transmission coefficient multiplied by exp(-$k_z$d), $k_z$ real] for a left-handed slab. For the perfect lens, the transfer function would be unity for all $k_x$. However, the deviation of either the real or imaginary part of µ limits the range of $k_x$, so that the slab acts as a low-pass filter. The losses, inherent to left-handed materials, also remove the singularities that appear in the transfer function.

**Figure 2:** Generic dispersion curve of $q_z$ versus $k_x$ for a negative index material having a periodic spatial modulation. For this example, a sinusoidal modulation has been assumed with period a=πc/2ω and modulation strength ∆=1.0. The solid curve corresponds to the real values of $q_z$, while the dashed line corresponds to the imaginary values of $q_z$.

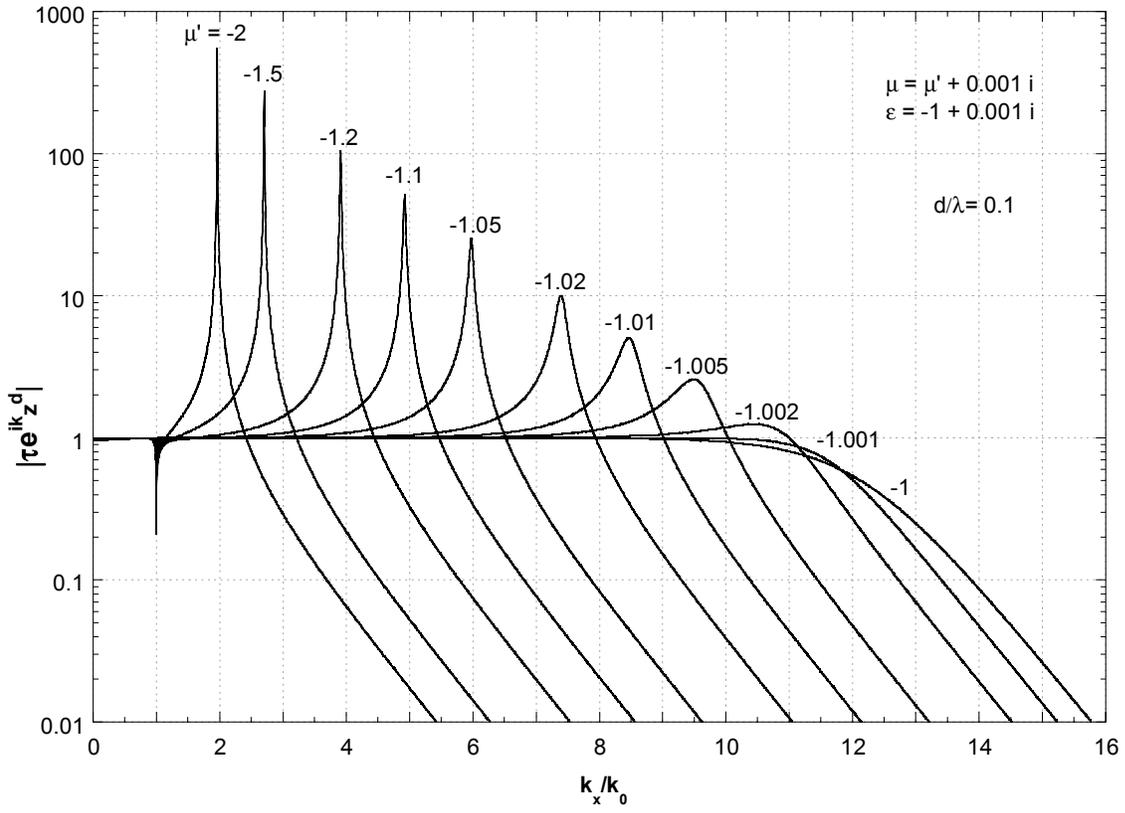

Figure 1

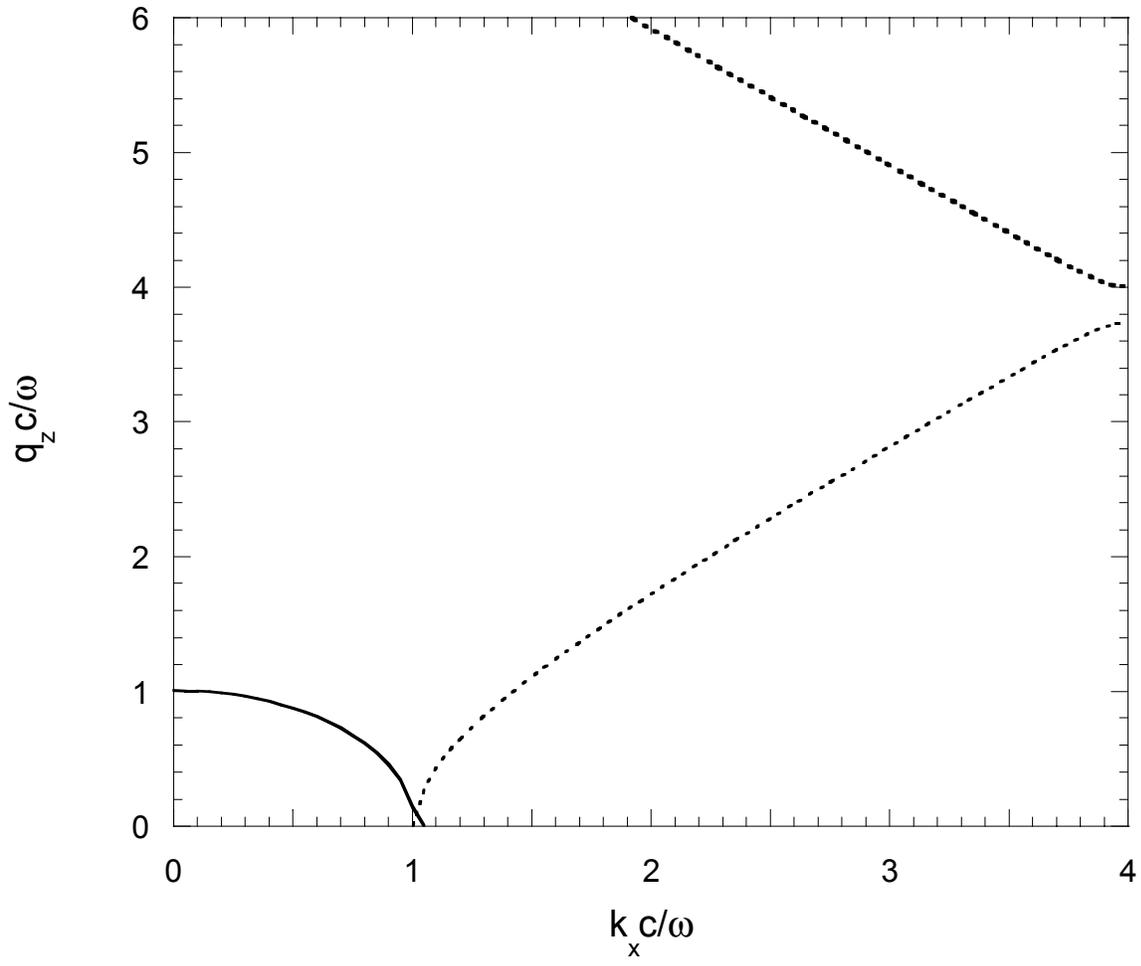

Figure 2